\documentclass[useAMS,usenatbib,usegraphicx]{mn2e}
\usepackage[dvips]{color}

\makeatother

\newcommand{\be}{\begin{equation}}
\newcommand{\ee}{\end{equation}}

\newcommand{\Msun}{M_{\odot}}

\definecolor{Black}{named}{Black}
\definecolor{Red}{named}{Red}

\title[Pinching of dark matter haloes]{Moderate Steepening of Galaxy Cluster 
Dark Matter Profiles by Baryonic Pinching}
\author[J.~Sommer-Larsen \& M.~Limousin]
{Jesper Sommer-Larsen$^{1,2}$\thanks{E-mail:jslarsen@astro.ku.dk (JSL); marceau.limousin@oamp.fr (ML)} \& Marceau
Limousin$^{3,1}$\footnotemark[1]\\
$^{1}$Dark Cosmology Centre, Niels Bohr Institute, University of Copenhagen,
  Juliane Maries Vej 30, DK-2100 Copenhagen, Denmark\\
$^{2}$Excellence Cluster Universe, Technische Universit\"at M\"unchen,
  Boltzmannstrasse 2, 85748 Garching, Germany\\ 
$^{3}$Laboratoire d'Astrophysique de Marseille, UMR\,6610, CNRS-Universit\'e de Provence,
   38 rue Fr\'ed\'eric Joliot-Curie, \\ 13\,388 Marseille Cedex 13, France}

\begin{document}

\date{Accepted 2009 August 31. Received 2009 August 30; 
in original form 2009 June 3}

\pagerange{\pageref{firstpage}--\pageref{lastpage}} \pubyear{2009}

\maketitle

\label{firstpage}

\begin{abstract}
To assess the effect of baryonic ``pinching'' of galaxy cluster dark matter
(DM) haloes, cosmological ($\Lambda$CDM) TreeSPH simulations of the formation 
and evolution of two galaxy clusters have been performed, with and without 
baryons included.

The simulations with baryons invoke
star formation, chemical evolution with non-instantaneous recycling, 
metallicity
dependent radiative cooling, strong star-burst, driven 
galactic
super-winds and the effects of a meta-galactic UV field, including
simplified radiative transfer.
The two clusters have $T$$\simeq 3$ and 6 keV, respectively, and,
at $z\sim0$, both host a prominent, central cD galaxy.

Comparing the simulations without and with baryons, it is found for the 
latter that
the inner DM density profiles, $r\la$ 50-100 kpc, steepen considerably:
$\Delta\alpha\sim$~0.5-0.6, where -$\alpha$ is the logarithmic DM density
gradient. This is primarily due to the central stellar cDs becoming very 
massive, as a consequence of the onset of late time cooling flows and 
related star formation. Once these spurious cooling flows have been
corrected for, and the cluster gravitational
potentials dynamically adjusted, much smaller pinching effects are found:
$\Delta\alpha\sim 0.1$. Including the effects of baryonic pinching,
central slopes of $\alpha\simeq1.0$ and 1.2 are found for the DM in the 
two clusters, interestingly close to recent observational findings for
galaxy cluster Abell~1703 based on strong gravitational lensing.

For the simulations with baryons, the inner density profile of DM and 
cluster gas 
(ICM) combined is found to be only very marginally steeper than that of the
DM, $\Delta\alpha\la0.05$. However, the total
matter inner density profiles are found to be $\Delta\alpha\sim0.5$ steeper 
than the inner profiles in the dark matter only simulations. 
       
\end{abstract}


\section{Introduction}
Large N-body cosmological simulations have been carried out for a decade, with the goal of making
statistical predictions on dark matter (DM) halo properties.
Because of numerical issues, most of these large cosmological simulations contain dark matter particles only.
They all reliably predict that the 3D density profile $\rho_{\rm DM}(r)$ should fall
as $r^{-3}$ at large radii (but within the virial radius). Observations have confirmed these predictions for cluster sized haloes
\citep[\emph{e.g.}][]{kneib03,etienne,rachel08,okabe09}.
This agreement is likely to be connected with the fact that at large radius, the density profile of a
galaxy cluster is dark matter dominated and the influence of baryons can be neglected.
On smaller scales, parameterizing the 3D density profile of the DM using a cuspy profile $\rho_{\rm DM}\propto
r^{-\alpha}$; dark matter only simulations predict a logarithmic slope $\alpha \sim$ 1-1.5 for
$r\,\rightarrow$ 0. The exact value of the central slope and its universality is debated \citep{nfw,moore98,ghigna00,
ricotti,navarro04,gao07}.
Theoretical efforts indicate that the inner logarithmic slope of DM halos 
should be $\alpha \sim$ 0.8 \citep{austin,steen}.
Although clearly of interest, these dark matter only studies and
their predictions
do not help much to make comparison with observations. Indeed, observing the central part
(i.e. the inner $\sim$ 500\,kpc) of a galaxy cluster at any wavelength reveals the presence of baryons (in the forms of stars and X-ray emitting hot gas).
Thus any attempt to compare observations to simulations in the center of galaxy clusters has to be
made with numerical simulations (or calculations) taking into account the baryonic component and its associated
physics.

On the numerical side, efforts are currently being made to include effects of
baryons in a realistic way in the simulations (see below). 
Problems remain, however: for example, due to the so-called ``over-cooling
problem'', numerical simulations tend to predict central brightest cluster galaxies which are too blue and too massive relative to observations.
Different effects do compete when it comes to the central slope of the density profile:
the cooling of gas in the central regions of galaxies and clusters is expected to lead to a more concentrated dark matter
distribution \citep[the so-called adiabatic contraction, see][]{blumenthal,gnedin,gustafsson}.
On the other hand, dynamical friction heating of massive galaxies against the diffuse cluster dark matter could in principle 
flatten the slope of the DM density profile 
\citep{elzant01,nipoti03,nipoti,MBK}.
Also, the properties of the inner part of simulated galaxy clusters (even in dark matter only simulations)
can depend significantly on initial conditions as demonstrated in \cite{maxwellhalo}.
Consequently, no coherent picture has yet emerged from N-body simulations when it comes to the shape
of the inner density profile of structures.

On the observational side, efforts have been put on probing the central slope $\alpha$ of the
underlying dark matter distribution. These analyzes have led to wide-ranging results, whatever the method
used: X-ray
\citep[][]{ettori02,arabadjis02,lewis03,zappacosta06};
lensing \citep{tyson98,S01,dahle03,sand02,gavazzi03,gavazzi05,sand04,sand07,bradac,L.08,R.09,O.09} or dynamics 
\citep{kelson02,bivianosalucci}.
This highlight the difficulty of such studies and the possible large scatter in the value of $\alpha$ from one cluster to another.

In summary, one needs to probe observationally and numerically the behavior of the \emph{underlying} dark matter 
distribution (i.e. after the baryonic component has been separated from the
dark matter component) in the central parts of galaxy clusters. 
The main difficulties are: i) Observationally, to be able to disentangle the baryonic component and the
underlying dark matter distribution;
ii) Numerically, to implement the baryonic physics into the simulations;
iii) Then to compare both approaches in a consistent way.

\cite{sand02,sand04,sand07} carried out lensing analyzes aiming to probe the central mass 
distribution in six galaxy clusters, using the measured velocity dispersion profile
of the cD galaxy as an extra constrain. They found central density slopes smaller than 1.

Recently, there have been a growing interest on Abell~1703,
a massive $z=0.28$ \citep{allen92} X-ray cluster with a luminosity
L$_{\rm{x}}$ = 8.7$\times10^{44}$ erg\,s$^{-1}$ \citep{hans00}.  
It contains a large number of gravitational arcs, enabling a very
detailed lensing analysis. Moreover,
although it displays an intriguing filamentary structure along the north-south direction,
it looks rather circular from the geometrical configuration of its multiply imaged systems,
in particular its giant arc, located at large angular separation ($\sim35\arcsec\sim147$\,kpc).
It is a uni-modal cluster, likely to be relaxed and characterized by a central 
dominant elliptical cD galaxy. This makes it much easier to interpret 
the results of the modeling compared to bimodal clusters such as Abell~1689 \citep{mypaperIII,RS.09}, Abell~2218
\citep{ardis2218}, Abell~68 \citep{a68} or MS\,2053.7-0449 (Verdugo et~al., 2007).
In fact, regular relaxed clusters are rare at such redshifts \citep{smith05}.
Finally, it displays a remarkable lensing configuration, forming a central
``ring'' composed of four bright images that represent a rare example of lensing by a hyperbolic
umbilic catastrophe \citep{orban}.
This ring is located close to the central cD galaxy (4-9 $\arcsec$, corresponding to 
17-38\,kpc), providing a robust constraint in the very central part of the cluster.

This motivated several lensing works on Abell~1703 aimed to probe the dark matter density slope of the 
central region ($\sim 20-200$\,kpc),
modeling the dark matter distribution of Abell~1703 by a
generalized NFW profile, viz.
\begin{equation}
\rho(r)=
\frac{\rho_{c} \delta_{c}}
{(r/r_{s})^{\alpha_{NFW}}(1+(r/r_{s}))^{3-\alpha_{NFW}}},
\end{equation}
where $r_{s}$ is the scale radius.
These works \citep{L.08,saha09,R.09,O.09} determined the inner
slope $\alpha_{NFW}$ to be consistent with the universal profile ($\alpha_{NFW}=1$).
The more recent strong lensing analysis by \cite{R.09}, based on the identification of 16 
multiply imaged systems,
of which 10 are spectroscopically confirmed, found $\alpha_{NFW}$=0.92$\pm$0.04 (1$\sigma$ confidence level).
Note that the scale radius derived by \cite{R.09} is found in agreement with that found by
\cite{O.09} when combining strong and weak lensing. Since the scale radius needs weak lensing
data to be properly constrained, this is important given the degeneracies arising between the scale radius and
the inner slope.

Assuming that pure dark matter haloes are well described by the standard
NFW profile with $\alpha_{NFW}$=1, the indication of the above results on Abell~1703 is that
the baryonic pinching of the dark matter halo, probably mainly caused
by the central dominant cD galaxy, does not substantially affect the
dark matter density slope in the central region of the cluster.   

On the theoretical/numerical front it has only recently been possible to 
carry out fully cosmological 
gas-dynamical/N-body simulations of the formation and evolution of galaxy
clusters at a sufficient level of numerical resolution and physical 
sophistication that the cool-out of gas, star-formation,
chemical evolution and gas inflows and outflows related to
individual cluster galaxies can be modeled to a reasonable degree of 
realism \citep[e.g.,][]{K.02,T.02,V03,T.04,S.05,D.05,R.05,R.06,Sa.06,M.06,
T.07,R.08}. 

\cite{R.05}, \cite{R.06}, \cite{D.05} and \cite{S.05} presented fully
cosmological simulations of galaxy groups and clusters. The TreeSPH
code used was building on the code used for simulating galaxy formation 
\citep[e.g.,][]{SL.03}, improved to include modeling of non-instantaneous 
chemical evolution \citep{L.02}, metallicity-dependent, atomic radiative 
cooling, strong supernova, and (optionally) AGN, driven galactic winds 
and thermal conduction. The two clusters simulated have $z$=0 virial
masses $M_{vir}\sim3\times10^{14}$ and $1.2\times10^{15} \Msun$,
one approximately the size of the Virgo cluster and the other of the 
Coma cluster. They were both selected
to be fairly relaxed, and both display central prominent cD
galaxies at $z$=0 as well as at $z$=0.28. In this paper we reanalyze
the simulations of the two clusters with emphasis on the dark matter
density profiles. We also present re-simulations of the clusters aimed at
correcting for the effects of spurious, late time cooling-flows.
Finally, we present results of new simulations of
the {\it same} two clusters, identical to the previous ones except that they
were carried out with dark matter only. We then compare the results
of simulations with and without baryons, to quantify the effects of baryons
(especially the cDs) on galaxy cluster dark matter profiles.   

The paper is organized as follows:
the code and the simulations are described in section 2, the results
obtained are presented in section 3 and discussed in section 4, and, 
finally, section 5 constitutes the conclusion.

Throughout the paper a
$\Lambda$CDM cosmology with $\Omega_{\rm{M}} = 0.3, \
\Omega_\Lambda = 0.7$ and a Hubble constant \textsc{H}$_0 = 70$
km\,s$^{-1}$ Mpc$^{-1}$ is assumed. 

\begin{table*}
\caption{Numerical and physical  
characteristics of the cluster simulations: mass of DM/gas/star particles and the respective gravitational (spline) softening
lenghts; total number of particles and initial redshift of each run; virial mass, radius and
temperature of simulated clusters (last three columns, referring to $z$=0).}
\begin{tabular}{l c c c c c c c c c c c}
\hline
run &   $m_{DM}$ &  $m_{gas}$ & $m_{*}$ & $\epsilon_{DM}$  &  $\epsilon_{gas}$  &  $\epsilon_{*}$  &
$N_{tot}$  &  $z_i$ &  $M_{vir}$  &  $R_{vir}$  &  $kT$ \\
        &  & [$10^7 M_{\odot}/h$] &  &  & [kpc/$h$] &  & & & [$10^{14}~M_{\odot}$] &
        [Mpc] &    [keV] \\
\hline  
Coma      & 180 & 25 & 25 & 5.4 & 2.8 & 2.8 & 950000 & 19 & 12.4 & 2.9 & 6 \\
Virgo   & 23 & 3.1 & 3.1 & 2.7 & 1.4 & 1.4 & 2235000 & 39 & 2.8 & 1.7 & 3 \\
VirgoLR      & 180 & 25 & 25 & 5.4 & 2.8 & 2.8 & 260000 & 19 & 2.8 & 1.7 & 3 \\
ComaDM   & 26 & - & - & 2.8 & - & - & 4000000 & 39 & 12.4 & 2.9 & - \\
VirgoDM & 26 & - & - & 2.8 & - & - & 1200000 & 39 & 2.8 & 1.7 & - \\
\hline
\end{tabular}
\label{tab:data}
\end{table*}

\section{The code and simulations}
The code used for the simulations is a significantly improved version of
the TreeSPH code we have used previously for galaxy formation simulations 
\citep{SL.03}: Full details on the code are given in \cite{R.06}; here we
recall the main improvements over the previous version.
(1) In lower resolution regions (which will always be
present in cosmological CDM simulations) an improvement in the numerical
accuracy of the integration of the basic equations is obtained by
solving the entropy equation rather than the thermal energy equation ---
we have adopted the ``conservative'' entropy 
equation solving scheme suggested
by \cite{SH02}. 
(2) Cold high-density gas is turned into stars in a probabilistic way as
described in \cite{SL.03}. In a star-formation event an SPH particle
is converted fully into a star particle. Non-instantaneous recycling of
gas and heavy elements is described through probabilistic ``decay'' of star 
particles back to SPH particles as discussed by \cite{L.02}. 
In a decay event a star particle is converted fully into an SPH particle.
(3) Non-instantaneous chemical evolution tracing
10 elements (H, He, C, N, O, Mg, Si, S, Ca and Fe) has been incorporated
in the code following Lia et~al.\ (2002a,b); the algorithm includes 
supernov\ae\ of type II and type Ia, and mass loss from stars of all masses.
Metal diffusion in Lia et al.\ was 
included with a diffusion coefficient $\kappa$ derived from models of the 
expansion of individual supernova 
remnants. A much more important effect in the present simulations, however,
is the redistribution of metals (and gas) by means of star-burst driven
``galactic super--winds'' (see point 5). This is handled self-consistently by
the code, so we set $\kappa$=0 in the present simulations. 
(4) Atomic radiative cooling depending both on the metal abundance
of the gas and on the meta--galactic UV field, modeled after Haardt
\& Madau (1996) is invoked. We also include a simplified treatment
of radiative transfer, by switching off the UV field where the gas
becomes optically thick to Lyman limit photons on scales of $\sim$ 1~kpc.
(5) 
Star-burst driven, galactic super-winds are incorporated in the simulations.
This is required to expel metals from the galaxies and get the 
abundance of the ICM to the observed level of about 1/3 solar in iron. 
A burst of star formation is modeled in the same way as the ``early 
bursts'' of \cite{SL.03}, i.e.\ by halting cooling in the surrounding gas 
particles, to mimic the initial heating and subsequent adiabatic expansion 
phase of the super-shell powered
by the star-burst; this scheme ensures effective energy
coupling and feedback between the bursting star particle and the surrounding 
gas. The strength of the super-winds is modeled
through a free parameter $f_{\rm{wind}}$ which determines how large a fraction
of the new--born stars partake in such bursting, super-wind 
driving star formation. We find that
in order to get an iron abundance in the ICM comparable to observations,
$f_{\rm{wind}}\ga0.5$ and at the same time a fairly top-heavy Initial Mass
Function (IMF) has to be used. 
(6) 
Thermal conduction was implemented in the code following \cite{CM99}.  

In this paper we present results for two simulated clusters,
``Virgo'' and ``Coma''. Both systems were chosen to be 
fairly relaxed (no $\ga$1:2 merging at $z\la$1). Virial masses at $z$=0 
are approximately 2.0x10$^{14}$ and 8.7x10$^{14}$ $h^{-1}$M$_{\odot}$ 
and X-ray emission weighted temperatures 3.0 and 6.0 keV, respectively. 
The clusters were selected from a
cosmological, DM-only simulation of a flat $\Lambda$CDM model, with
$\Omega_M$=0.3, $\Omega_b$=0.036, $h$=0.7 and $\sigma_8$=0.9 and a box-length
of 150 $h^{-1}$Mpc. 
Mass and force resolution was increased in, and gas particles added to,
Lagrangian regions enclosing the clusters. The Virgo cluster was then
re-simulated using 2.3 million baryonic+DM particles with $m_{\rm{gas}}$=$m_*$=
3.1x10$^7$ and $m_{\rm{DM}}$=2.3x10$^8$ $h^{-1}$M$_{\odot}$ for the high 
resolution gas, star and dark matter particles. Gravitational (spline) 
softening lengths of $\epsilon_{\rm{gas}}$=$\epsilon_*$=1.4 and 
$\epsilon_{\rm{DM}}$=2.7 $h^{-1}$kpc, respectively, were adopted.
The more massive Coma cluster was re-simulated using 1.0 million baryonic+DM 
particles with $m_{\rm{gas}}$=$m_*$=
2.5x10$^8$ and $m_{\rm{DM}}$=1.8x10$^9$ $h^{-1}$M$_{\odot}$, 
$\epsilon_{\rm{gas}}$=$\epsilon_*$=2.8 and 
$\epsilon_{\rm{DM}}$=5.4 $h^{-1}$kpc. As a resolution test, the Virgo
cluster was also simulated at this (eight times lower) mass and 
(two times lower) force resolution. This simulation will be denoted
``VirgoLR''.

For all simulations, gravitational softening lengths were fixed in co-moving
coordinates till $z$=6, subsequently in physical coordinates.

For the simulation presented in this paper $f_{\rm{wind}}$=0.8,
and an Arimoto-Yoshii IMF (of slope $x$=--1, shallower than the 
Salpeter slope $x$=-1.35) with mass limits [0.1--100]~$\Msun$
was adopted. AGN driven winds were not invoked.
Moreover, the thermal 
conductivity was set to zero assuming that thermal conduction in the ICM is 
highly suppressed by magnetic fields \citep[e.g.,][]{EF00}. We note, that
in relation to all relevant quantities presented in 
this paper, no significant difference is found between
simulations with zero thermal conductivity and simulations invoking thermal
conduction at 1/3 of the Spitzer level \citep{R.05}.

\subsection{Correcting for the effects of late quasi-stationary cooling flows}
After a period of major merging at $z \ga 2$ strong,
quasi-stationary cooling flows develop at the centers of the clusters despite 
the strong, super-nova driven energy feedback to
the IGM/ICM through galactic super-winds and the use of a fairly top-heavy
IMF. As a consequence, stars continue to form steadily at the centers of the
cDs ($r\la$10 kpc) at rates of $\sim$100 and $\sim$500 $M_{\odot}$/yr at $z$=0
for Virgo and Coma, respectively. As discussed in \cite{S.05}, this results in 
the cDs becoming
too blue over time, with central B-R colours of $\sim$1.0 at $z$=0, 
rather than 1.4-1.5 observed for real cDs (note though that some 
star-formation is observed in many cDs at the base of the cooling 
flow --- e.g., McNamara 2004). Given the significant gas cool-out rates,
the increase in the stellar masses of the cDs since $z\sim$2 is considerable:
For the Virgo cD $M_{*,12}$=0.42, 1.0 and 1.4 in units of $10^{12} \Msun$,
at $z$=2,1,0, respectively (inside of 20 kpc). 
For Coma, the corresponding masses are
0.95, 1.9 and 4.2. In real cDs, the late time cooling flows will be
counteracted, most likely by the energy release related to the accretion
of mass onto super-massive black holes \citep[e.g.,][]{B.06,C.06}.
This will lead to a strong suppression
of the late time star formation rate, such that the simulated cDs become
too massive. As will be shown in the next section, the gravitational
force of this excess mass influences the central dark matter profile, 
and must be corrected for.

Assuming that the excess mass has been added adiabatically, i.e. on a
timescale large compared to the stellar orbital timescale in the cD,
the ``correct'' dark matter distribution can be recovered by slowly
removing the excess stellar mass. This is accomplished as follows:

In order to enable a direct comparison to the observational results
obtained for Abell~1703 we focus on dark matter profiles at $z$=0.28,
corresponding to $t$=10.2 Gyr. At $t$=8.2 Gyr, stars at $r<20$ kpc
from the center of the cD are identified. For each of the two simulated
clusters two additional simulations are carried out --- in one the
masses of all identified stars formed since $z$=1 are then gradually reduced to
zero over a period of 1 Gyr, in the other the masses of all identified
stars formed since $z$=2.
At the same time radiative cooling as well as star formation is switched
off. Subsequently the additional simulations are then continued for 
2 Gyrs more, still with radiative cooling and star formation switched off.
We refer to these additional simulations as Rz1 and Rz2 respectively.

The purpose of these additional simulations is to, without resorting
to completely new simulations with AGN feedback etc. included, determine 
the cluster dark matter profiles in the (more realistic) case where
effects of late time cooling flows are counteracted since redshifts
of either 1 or 2. 

\subsection{Dark matter only simulations}
To further compliment the existing simulations and determine the role
of the inclusion of baryonic physics, we carried out dark matter only
simulations of the two clusters using the same initial
conditions as previously, but not splitting the particles into SPH and
DM particles. 

The Virgo and Coma cluster were re-simulated using 1.2 and 4.0 million 
DM particles, respectively. These simulations will be denoted ``VirgoDM''
and ``ComaDM'' in the following. The high resolution DM particles had
$m_{\rm{DM}}$=2.6x10$^8$ $h^{-1}$M$_{\odot}$ and  
$\epsilon_{\rm{DM}}$=2.8 $h^{-1}$kpc.\\[1cm]
Numerical and physical parameters for the simulations and the
$z$=0 clusters are summarized in Table.~\ref{tab:data} 

\begin{figure}
\includegraphics[width=84mm]{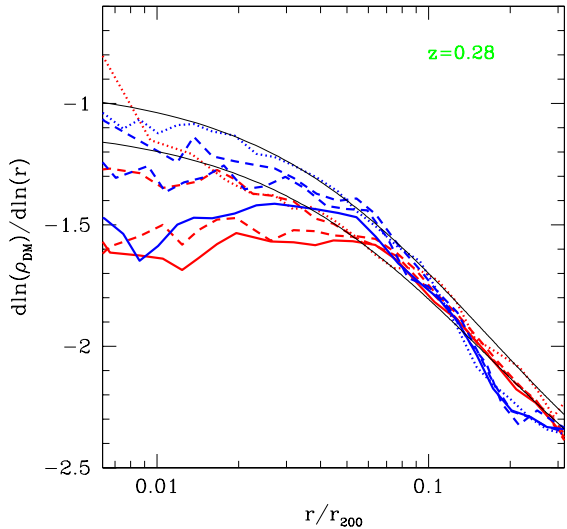}
 \caption{Logarithmic density slopes of the spherically averaged
   cluster dark matter distributions at $z$=0.28. Solid red curve shows result
   for the ``Virgo'' cluster including the cD uncorrected for effects of
   late time cooling flows. Dotted red curve shows the result of the
   Virgo DM only simulation. The two red dashed lines show the results
   of the Rz1 (lower) and Rz2 (upper) re-simulations, where effects
   of spurious late time cooling flows have been corrected for. The
   blue curves show the same for the ``Coma'' cluster. The two black
   solid curves show generalized NFW models (eq.[1]) with
   $\alpha_{NFW}$=0.92 (upper) and 1.09 (lower), and concentration parameter 
   $c$=$r_{200}/r_s$=6.
}
\label{fig:dlnrhodlnr}
\end{figure}

\section{Results}
\subsection{Inner dark matter halo profiles}
To enable a precise determination of the galaxy cluster dark matter 
density profiles at $z$=0.28, 21 time frames, with a time spacing of 
0.1 Gyr (large enough to enable quasi-random sampling of the central
halo profile) and centered on $t$=10.2 Gyr, were co-added. The DM
profile power law index, 
\begin{equation}
\alpha(r) = -\frac{dln(\rho_{\rm{DM}}(r))}{dln(r)} ~~,
\end{equation}
where $r$ is the cluster-centric radial distance, was then determined 
by averaging the co-added DM distribution over
spherical shells. 

In Fig.~\ref{fig:dlnrhodlnr} is shown by dotted lines
the resulting $\alpha(r)$ (strictly speaking -$\alpha(r)$, but
we shall neglect this distinction in the following) for the dark matter only simulations.
Moreover, $\alpha(r)$ is shown by thick solid curves for the original 
simulations including baryons, i.e. simulations including baryons, but 
without any corrections for late cooling-flows and related star formation.
Finally, shown by thin solid lines, is $\alpha(r)$ for 
generalized NFW profiles (eq.[1]) 
with $\alpha_{NFW}$=$\alpha(0)$=1.09 and 0.92, as determined for Abell 1703 by 
\cite{L.08} and \cite{R.09}, respectively, and concentration parameter 
$c$=6. It is seen
from the figure that a) all the simulations match the NFW profiles 
in the intermediate and outer parts of the haloes, $r/r_{200}\ga$~0.07,
fairly well, b) the dark matter only simulation of Virgo is matched
quite well by an $\alpha(0)\simeq1.1$ NFW profile and that of Coma
by an $\alpha(0)\simeq0.9$ profile, and c) the simulations with baryons
result
in too large values of $\alpha(r)$ ($\sim1.5-1.6$) in the inner parts, 
$r/r_{200}\la$~0.04 (the results are affected by effects of gravitational
softening at $r/r_{200}\la$~0.007-0.008 --- see section 3.2.1. for more
detail).  In order to assess whether this is due to
baryonic ``over''-pinching, i.e. due to the simulated cDs being
unrealistically massive, we compare the simulated and observed Virgo
and Coma clusters: 

At $z$=0 the cDs in the simulated clusters
have masses (inside of $r$=50 kpc --- see below) of 7.4$\times10^{12}$ and 
2.2$\times10^{12} \Msun$, for Coma and Virgo, respectively (at
$z$=0.28 the corresponding numbers are 5.1$\times10^{12}$ and
1.8$\times10^{12} \Msun$). 
The real Coma and Virgo clusters both contain two dominant galaxies:
NGC~4874 and NGC~4889 in Coma, and M86 and M87 in Virgo. The two
dominant galaxies in each cluster will eventually merge into one
cD in each.
The B-band luminosities of NGC~4874 and NGC~4889 are
1.5$\times10^{11}$ and 1.8$\times10^{11} L_{\odot,B}$, respectively 
\citep{MA08}. 
Assuming a B-band M/L of about eight (as found for the cDs in the
present simulations based on the Arimoto-Yoshii IMF; the
more standard Salpeter IMF results in similar M/L) the combined stellar mass
of the two galaxies (at $z \simeq 0$) is 2.6$\times10^{12} \Msun$.
Moreover, the simulated Coma cluster is a $T_X\sim$6 keV cluster, whereas
the real Coma cluster is a $T_X\sim$8 keV cluster. Assuming a standard scaling
of mass $\propto T_X^{3/2}$ the cD in a 6 keV cluster should have a stellar
mass of 1.7$\times10^{12} \Msun$.
The B-band luminosities of M86 and M87 are
4.2$\times10^{10}$ and 6.6$\times10^{10} L_{\odot,B}$, respectively 
\citep{B.85}. The stellar masses of the two galaxies thus add up to
about 8.6$\times10^{11} \Msun$.

The simulated cDs are hence about
3-4 times more massive than observed, indicating that the simulated
DM haloes are too contracted compared to real ones, which at least 
qualitatively could explain some of the differences seen in 
Fig.~\ref{fig:dlnrhodlnr}.

To understand the differences seen in Fig.~\ref{fig:dlnrhodlnr} more
quantitatively we now turn to the re-simulations aimed at correcting
for the effects of the late cooling flows. In Fig.~\ref{fig:dlnrhodlnr} 
we show by thick dashed lines $\alpha(r)$ for the simulations with baryons, 
where the mass of all stars in the central cD, formed since either 
$z_f$=1 or 2, has been adiabatically
reduced to zero --- we shall denote these two sets of re-simulations
by Rz1 and Rz2 in the following.
For the inner parts of Coma, a large effect on
$\alpha(r)$ results already from removing central stars formed since $z_f$=1.
This is related to the fact that the mass of the cD more than doubles
since $z$=1 in the original simulation including baryons. For Virgo
a strong effect is seen only between the original simulation and $z_f$=2 
correction. This is also reasonable, since the mass of the cD more than
doubles going from $z$=2 to 1, but then only increases by additional 40\%
going to redshift 0. In order to assess which
of the Rz1 and Rz2 re-simulations, if any, result in cDs
which are realistic, we now derive the properties of the resulting
cDs.

\subsection{Properties of the simulated cD galaxies}

\begin{figure}
\includegraphics[width=84mm]{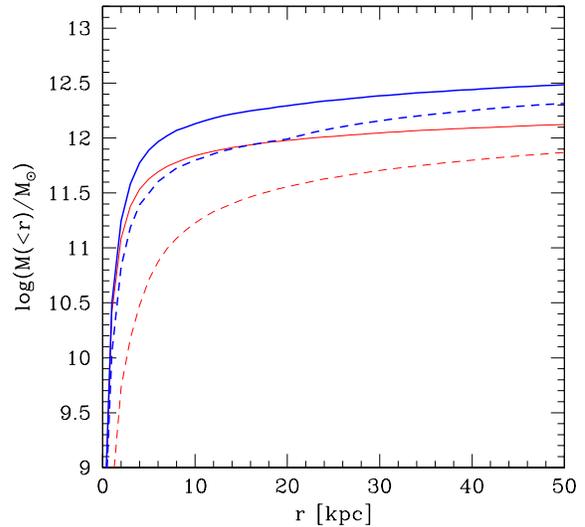}
 \caption{For re-simulations Rz1 (solid curves) and Rz2 (dashed curves), 
   at $z$=0, are
   shown the cD cumulative stellar masses (Virgo: red curves; Coma: 
   blue curves).}   
\label{fig:mcum}
\end{figure}

\subsubsection{cD stellar masses}
In Fig.~\ref{fig:mcum} is shown the cumulative stellar mass for the
Coma and Virgo cDs from the $z_f$=1 and 2 adiabatically corrected
simulations, respectively. It is seen, that at $r$=50 kpc, corresponding
to approximately the 25 B-mag/arcsec$^2$ (optical) isophote (see below),
for $z_f$=1 the cumulative stellar masses are 3.1 and 
1.3$\times10^{12} \Msun$ for Coma and Virgo, respectively. For $z_f$=2 the
corresponding numbers are 2.1 and 0.74$\times10^{12} \Msun$. Comparison to
the observational estimates above, indicates that the $z_f$=2 cDs provide
a better match to reality than the $z_f$=1 ones.

An alternative comparison can be made to the cD of Abell~1703, for
which \cite{L.08} estimate a stellar mass of 1.3$\pm$0.3 $\Msun$ inside
of $r$=30 kpc. The temperature of the Abell~1703 ICM is
uncertain, but (also uncertain) estimates of the virial mass of
Abell~1703 indicate that the temperature is likely at least as large as 
for the simulated Coma cluster (Riemer-S{\o}rensen, private communication,
Oguri et~al 2009). For the Coma Rz1
and Rz2 cDs, at $z$=0.28, we find stellar masses inside of $r$=30 kpc
of 2.2 and 1.2$\times10^{12} \Msun$. Again, the stellar mass
appears realistic for the Rz2 Coma cD, but too large for the Rz1 cD.       

\subsubsection{cD effective radii}
Next, we determine the effective radii for the cDs at $z$=0. 
To this end, the projected B-band surface brightness profiles 
are determined for the cDs (see Romeo et~al. 2005 for details on calculating
broad band photometric properties of the star particles):
First, a B-band surface brightness map is constructed by projecting
the star-particle distribution within a cube of side-length $L$=200 kpc, 
centered on the cD, along the three cardinal directions and then superposing
the three maps (larger values of $L$ result in similar results). 
Second, a radial surface profile is constructed by
azimuthally averaging over the combined image. Third, the radial
profile obtained is divided by a factor of three.
At $R\simeq$50 kpc the B-band surface 
brightness for both cDs drops to 25 mag/arcsec$^2$ \citep{S.05}, which
is taken to be the radial extent of the cD itself. $R_{\rm{eff}}$ is
then determined as the projected radius enclosing half of the cD B-band
luminosity. For the Coma and Virgo cDs in the Rz1 simulations,
$R_{\rm{eff}}$=8 and 7 kpc is found. The corresponding values for
the Rz2 simulations are $R_{\rm{eff}}$=16 and 11 kpc. For a large 
(observational)
sample of SDSS galaxies (at $z\sim$0), 
\cite{Sh.03} find for early type galaxies
of stellar masses 3.1 and 1.3$\times10^{12} \Msun$ median effective radii
of $\sim$25 and 16 kpc, and for stellar masses of 2.1 and 
0.74$\times10^{12} \Msun$, $R_{\rm{eff}}\sim$20 and 13 kpc. It follows
that the Rz1 cDs have too small effective radii for their mass, whereas
the Rz2 cDs have reasonable effective radii compared to observations.

\subsubsection{cD metal abundances and alpha/Fe ratios}
\begin{figure}
\includegraphics[width=84mm]{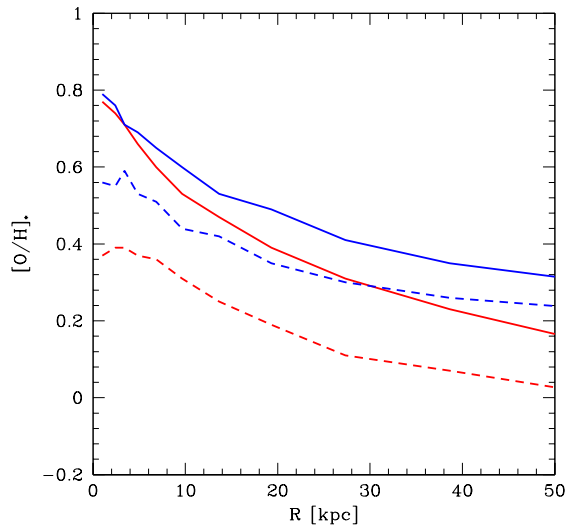}
 \caption{Stellar oxygen abundance as a function of projected radial
   distance from the center of the cD for re-simulations Rz1 and Rz2,
   at $z$=0. The curves are labeled as in Fig.~\ref{fig:mcum}.}
\label{fig:oh}
\end{figure}

In Fig.~\ref{fig:oh} is shown, for cDs in the Rz1 and Rz2 re-simulations, 
the projected average stellar oxygen abundance as a function of
projected radius $R$. In these alpha-element enhanced systems
(see below),
[O/H] can be taken as a measure of the total abundance [Z/H]. It is
seen that the oxygen abundance is super-solar all the way to 50 kpc
projected radius. For the Rz1 simulations, the integrated (global) cD 
abundances
are [O/H]=0.59 for both clusters. For the Rz2 simulations, the integrated
abundances are [O/H]=0.40 and 0.26 for Coma and Virgo, respectively.
Following \cite{T.05}, observed early type galaxies
of stellar masses 3.1 and 1.3$\times10^{12} \Msun$ have 
median $<$[Z/H]$>$=0.38 and 0.33, respectively, and 
for stellar masses of 2.1 and 0.74$\times10^{12} \Msun$, $<$[Z/H]$>$=0.36 and
0.30. So the Rz1 cDs have too large abundances compared to what is
observed, whereas the Rz2 cDs provide a good match to observations.
Taking [O/Fe] as a proxy for alpha/Fe, the results for the Rz1 and
Rz2 simulations are similar, [O/Fe]$\sim$0.3-0.35. These values
are within the range observed for large early type galaxies \citep{T.05}.

\begin{figure}
\includegraphics[width=84mm]{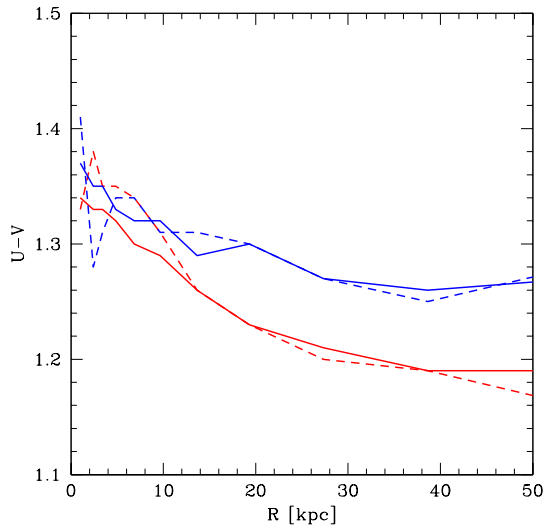}
 \caption{U-V colour as a function of projected radial
   distance from the center of the cD for re-simulations Rz1 and Rz2,
   at $z$=0. The curves are labeled as in Fig.~\ref{fig:mcum}.}
\label{fig:uv}
\end{figure}

\subsubsection{cD U-V colours}
For $z\sim0$ early type galaxies, the rest-frame U-V broadband colour allows
sampling of the 4000~{\AA} break, and therefore is particularly sensitive to
age and metallicity variations of the stellar populations.  
In order to check whether the simulated cDs fall on the observational
``red sequence'' for early type galaxies, we hence calculate the U-V 
colours for the cDs.
In Fig.~\ref{fig:uv} is shown, for cDs in the Rz1 and Rz2 re-simulations, 
the U-V colour as a function of projected radius $R$. In the inner parts
of the galaxies, all the cDs display a negative colour gradient, qualitatively
consistent with what is observed for large early type galaxies. As the
mean stellar age is approximately constant with projected radius $R$,
the reason for the gradient in colour is the metallicity gradient
seen in Fig.~\ref{fig:oh}. This is also the generally accepted
explanation for the observed colour gradients in early type systems
\citep[\emph{e.g.}][]{T.00}.
For the Rz1 cDs, integrated (global) U-V colours
of 1.31 and 1.27 are found for Coma and Virgo, respectively. For the 
R2z cDs, the corresponding numbers are U-V=1.29 and 1.28. The two Rz1 cDs have
$M_V$=-24.6 and -23.7 for Coma and Virgo, respectively. For the Rz2
cDs, the corresponding numbers are -23.8 and -23.0. Comparing to the
observational U-V vs. $M_V$ sequence \citep[\emph{e.g.}][]{R.08}, early
type galaxies of such absolute V magnitudes have a median U-V$\sim$1.4.
We hence find that the Rz1 and Rz2 cDs have very similar U-V colours,
and that these are slightly bluer than the observed average. This is,
at least partly, due to the large, somewhat bluer stellar envelope 
surrounding the cDs, cf. Fig.~\ref{fig:uv}.  
 
\subsubsection{cD central physical stellar densities}
From the results presented above it is possible to estimate the stellar
core density of the cDs, defined as
\begin{equation}
\rho_{c,*} = \frac{M_*(r<R_{\rm{eff}})}{4\pi/3~R_{\rm{eff}}^3} ~~.
\end{equation}
For the Rz1 cDs we find $\rho_{c,*}$=6 and 4$\times10^{8}$ $\Msun$/kpc$^3$
for Coma and Virgo, respectively. For the Rz2 cDs, the corresponding 
numbers are 5 and 4$\times10^{7}$ $\Msun$/kpc$^3$. Extrapolating the
observational estimates of $\rho_{c,*}$, given in \cite{vD.08} on the
basis of SDSS data, to galaxy stellar masses of $\sim10^{12}$~$\Msun$, the Rz2
cD core densities appear the most reasonable, although no strong
conclusions can be made on the basis of the available data.   

\subsection{How much are galaxy cluster dark matter haloes pinched by
the central cDs?}
Based on the results given in the previous subsection, it is clear
that the cDs formed in the Rz2 re-simulations give a much better match
to reality than the ones formed in the Rz1 simulations, and in fact,
in general, meet all observational constraints. We shall hence in the
following focus on the Rz2 re-simulations, and the comparison of these
to the dark matter only simulations.

\subsubsection{Resolution limitations due to gravity softening}
The Coma dark matter only simulation has $\epsilon_{\rm{DM}}$=2.8 $h^{-1}$kpc, 
and the gravity force of the central dark matter particles is hence 
purely Newtonian
(i.e. un-softened) at $r\simeq$8 kpc. At $z$=0.28, $r_{200}$ for the Coma
cluster is 1820 kpc, which means that the simulation result is affected
by resolution effects inside of $r/r_{200}\simeq$0.004. The Coma simulation
with baryons has $\epsilon_{\rm{DM}}$=5.4 $h^{-1}$kpc, so although the
star and gas particle softening lengths are considerably smaller, the
resulting dark matter profile may be affected by gravity softening
inside of $r/r_{200}\simeq$0.008. The Virgo simulations with
and without baryons have $\epsilon_{\rm{DM}}$=2.7 and 2.8 $h^{-1}$kpc,
respectively. As $r_{200}$ for this cluster at $z$=0.28 is 1105 kpc,
the simulation results may be affected by gravity softening inside
of $r/r_{200}\simeq$0.007. Finally, the VirgoLR simulations have
$\epsilon_{\rm{DM}}$=5.4 and 5.6 $h^{-1}$kpc, respectively, resulting
in an inner radial resolution limit of $r/r_{200}\simeq$0.014.

In the following we shall only discuss
simulation results outside of the above gravity softening resolution
limits.

\begin{figure}
\includegraphics[width=84mm]{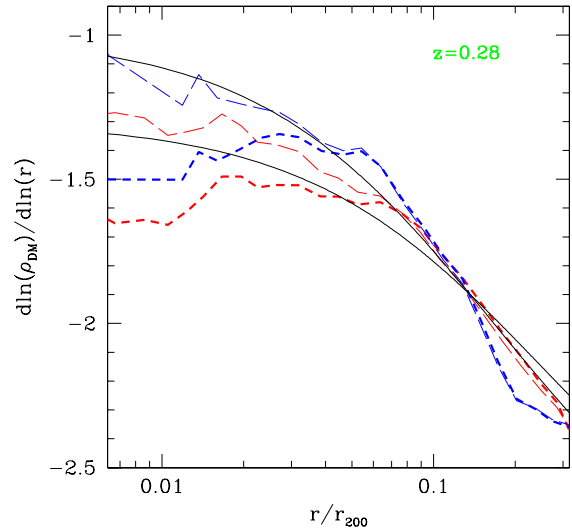}
\caption{Logarithmic density slopes for the Rz2 re-simulations. Results
   for total matter densities are shown by thick, short-dashed lines
   (red: Virgo; blue: Coma). The thin long-dashed lines shows the
   result for the dark matter (as in Fig.~\ref{fig:dlnrhodlnr}). 
   The two black solid curves show generalized NFW models (eq.[1]) with
   $\alpha_{NFW}$=1.0 and 1.3, and concentration parameters 
   $c$=$r_{200}/r_s$=4 and 6, respectively.}
\label{fig:dlnrhodlnr_all}
\end{figure}

\subsubsection{Steepening of the inner dark matter profiles}
As can be seen from Fig.~\ref{fig:dlnrhodlnr}, the dark matter density
profile obtained in the Coma dark matter only simulation, can be fairly
well described by a generalized NFW profile of $\alpha_{NFW}=\alpha(0)$$\sim$0.9.
Comparing it to what is obtained for the Rz2 re-simulation of Coma the
steepening of the inner dark matter profile, caused by the cD, is quite
moderate, $\Delta\alpha\sim$0.1. For the Virgo cluster, the dark matter
only profile is also fairly well described by a generalized NFW profile,
this time of $\alpha_{NFW}=\alpha(0)$$\sim$1.1. 
Again, the Rz2 re-simulation only leads to a very moderate steepening,
$\Delta\alpha\sim$0.1, relative to the dark matter only simulation. So
for realistic cDs, the steepening of the dark matter profile in galaxy
clusters is very moderate, $\Delta\alpha\sim$0.1. This is the most
important result presented in this work.

\subsubsection{Total matter density profiles}
As gravitational lensing probes the total matter (surface) density,
in order to deduce dark matter profiles observationally, one has to
do a dark matter vs. baryonic matter decomposition. This obviously
adds some uncertainty (statistical as well as possibly systematic)
to the estimate of the dark matter density profile. To bypass this,
it is clearly also of interest to calculate total matter density
profiles from the simulations. These can then more directly be
compared to observational finding, based on lensing. In 
Fig.~\ref{fig:dlnrhodlnr_all} is shown the result for the Rz2
simulations. Both the Virgo and the Coma profile are characterized
by a rather sharp break at $r/r_{200}\simeq$0.05-0.06. Inside of
this the profiles are rather flat, with $\alpha(r) \sim 
\alpha_{NFW;DMonly}+0.5$.
The profiles are not well described by generalized NFW models,
as exemplified in the figure.

\begin{figure}
\includegraphics[width=84mm]{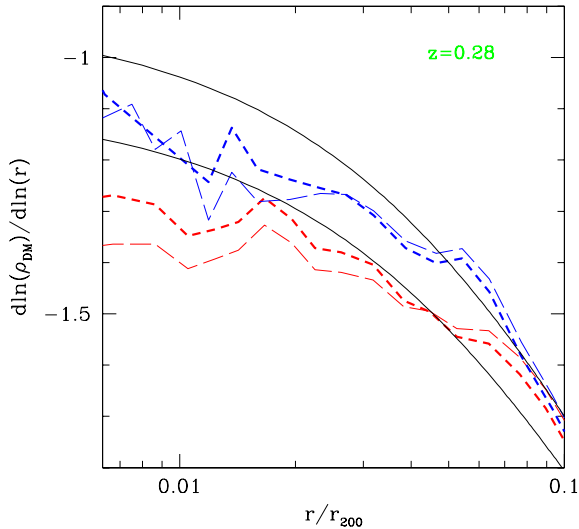}
 \caption{For the Rz2 re-simulations is shown $\alpha(r)$ for the 
   dark matter alone (thick short-dashed lines), and for dark matter and
   ICM gas combined (thin long-dashed lines). Results for Coma are
   shown by blue curves, for Virgo by red curves. The two black
   curves are the NFW models shown in Fig.~\ref{fig:dlnrhodlnr}.}
\label{fig:dlnrhodlnr_DM+gas}
\end{figure}

\begin{figure}
\includegraphics[width=84mm]{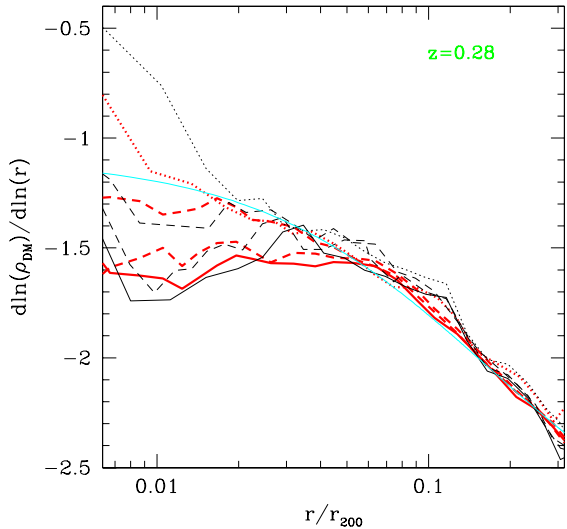}
 \caption{Comparison of the Virgo and VirgoLR simulations, at $z$=0. The 
   red curves display the results for Virgo, labeled in the same way
   as in Fig.~\ref{fig:dlnrhodlnr}. The black curves show the
   corresponding results for VirgoLR. The statistical uncertainty
   for the VirgoLR data is $\Delta\alpha\sim$0.05; for Virgo about a 
   factor of $\sqrt{8}$ less. The light blue curve shows
   a generalized NFW models (eq.[1]) with $\alpha_{NFW}$=1.09
   and concentration parameter $c$=$r_{200}/r_s$=6.}
\label{fig:res}
\end{figure}

\subsubsection{Dark matter + gas density profiles}
In performing the dark matter vs. baryonic matter decomposition mentioned
above, one typically assumes that the baryonic component is dominated
by the stellar mass of the cD, for which the
mass distribution can be fairly well modeled. It is hence important
to show, that neglecting the hot gas (ICM) component, for which the mass
distribution is much more uncertain, does not significantly bias
the estimate of the DM distribution.  
In Fig.~\ref{fig:dlnrhodlnr_DM+gas} is shown, for the Rz2 simulations,
$\alpha(r)$ for the dark matter alone, as well as for dark matter and
hot ICM gas combined. As can be seen, the effect on $\alpha(r)$ of 
including the ICM is very modest, $\Delta\alpha\la$0.05.  

\subsection{A numerical resolution test}
In Fig.~\ref{fig:res} is shown $\alpha(r)$ for the sets of Virgo simulations
as well as VirgoLR simulations. Statistical uncertainties on the
VirgoLR results are $\Delta\alpha\sim$0.05, and on the Virgo results
about a factor $\sqrt{8}$ less. Note that the VirgoLR results are
affected by gravity softening at about $r/r_{200}\simeq$0.014. Outside
of this region, there is reasonably good agreement between the two
sets of simulations, indicating that the
results presented in Fig.~\ref{fig:dlnrhodlnr} are not seriously affected
by numerical resolution limitations outside of $r/r_{200}\simeq$0.007-0.008. 

\section{Discussion}
Fitting the dark matter density profile of Abell~1703 by a generalized
NFW profile, different authors find 
$\alpha(0)$ between 0.8 and 1.2 \citep{L.08,saha09,R.09,O.09}.
Although we in this paper are analyzing simulations of just two
clusters, and although other observational values of $\alpha(0)$ have
previously been reported in the literature, it is obviously interesting
that we for the two clusters simulated, including the effects of
the central cD on the dark matter halo, find $\alpha(0)\simeq$1.0-1.2.

\begin{figure}
\includegraphics[width=84mm]{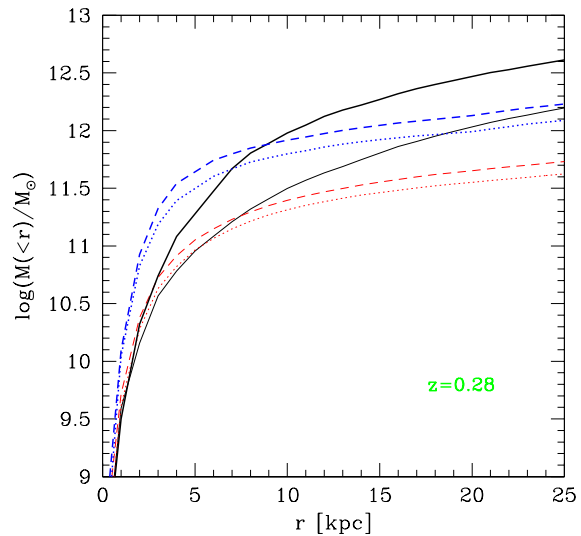}
 \caption{For the Rz2 runs, at $z$=0.28,  are shown cumulative mass of dark 
   matter (solid curves; Coma: thick, Virgo: thin), stars (dotted curves;
   Coma: blue, Virgo: red) and stars+gas (dashed curves; Coma: blue,
   Virgo: red).}
\label{fig:mcumall}
\end{figure}

Given that LCDM haloes undoubtedly have a range of inner slopes, and given
that we are probing only two haloes, an even more interesting result is
that for both clusters, the steepening of the inner slopes is both
moderate and consistent, $\Delta\alpha\sim$0.1. For Milky Way sized
galaxies, as well as smaller galaxies, \cite{gustafsson} found much
larger effects of baryonic pinching of the dark matter haloes, 
$\Delta\alpha\sim$0.6. To understand the reason for this difference,
we show in Fig.~\ref{fig:mcumall}, for the Rz2 re-simulations at
$z$=0.28, the cumulative masses of dark matter, stars and stars+gas.
As can be seen, the cumulative masses of baryons and dark matter are
equal at $r_{eq}\simeq$7 and 9 kpc, for Virgo and Coma respectively. In
both cases this is well inside of the effective radius, i.e. the 
characteristic scale of the galaxy. In contrast, \cite{gustafsson}
found $r_{eq}\sim$10 kpc for the individual (field) galaxies considered.
These galaxies have characteristic scales of $\sim$2-4 kpc, so the
chief difference is that the cD galaxies considered in this paper are
much less baryon dominated than normal (substantially smaller) 
field galaxies. In addition,
one should note that the present simulations can only probe the
increase in dark matter inner profile steepening outside of $\sim$8 
and 16 kpc, for Virgo and Coma respectively (whereas the innermost,
baryon dominated regions are about twice as well resolved due to
the shorter gravitational softening lengths of the star and gas
particles).

It is also worth noting that \cite{gnedin} found a consideably
larger steepning of galaxy cluster inner DM profiles due to
baryonic pinching, on the basis of their simulations invoking baryonic
physics. This, however (as acknowledged by these authors), is due to the 
spurious effects of ``over-cooling'', discussed and corrected for
in the present work.

An underlying assumption in the Rz1 and Rz2 re-simulations is that
the rate of change of cD gravitational potential, caused by stars 
either being formed {\it in situ} or being added through accretion
since $z_f$=1 or 2, is slow (adiabatic), compared to the dynamical 
rate. Although this is true for the stars forming {\it in situ} at
the base of the cooling flows, in the Coma simulation $\sim$1:3 and
1:4 mergers take place at $z\simeq$1.4 and $z\simeq$0.7, respectively, 
and in the Virgo simulation $\sim$1:4 mergers take place, at $z\simeq$1.5 
and 1.0, respectively.
In such mergers, the adiabatic approximation is not fulfilled, and
more realistic simulations, including effects of super-massive 
black hole growth and AGN feedback, would have to be undertaken to
check whether the results we find from the adiabatic stellar mass
removal re-simulations are in fact fully correct. However, the
fact that the Rz2 cDs appear to be realistic in terms of stellar
mass, linear extent, metallicity, etc. can be considered very strong
evidence that the inner dark matter profile steepening found for
the Rz2 simulations are entirely reasonable. It is obvious that
the primary drivers of the baryonic pinching are a) cD stellar mass,
and, to a lesser extent, b) the cD linear extent.  

\section{Conclusion and outlook}
We have in this paper presented results of cosmological ($\Lambda$CDM) 
TreeSPH simulations of the formation and evolution of two galaxy clusters.
In order to quantify the effect of baryonic ``pinching'' of galaxy cluster 
dark matter haloes, we have simulated {\it the same} clusters with and without 
baryons included.

The hydrodynamical simulations include a number of important physical 
processes, such as
star formation, chemical evolution with non-instantaneous recycling, 
metallicity
dependent radiative cooling, strong star-burst, driven 
galactic
super-winds and the effects of a meta-galactic UV field, including
simplified radiative transfer.
The two clusters have $T$$\simeq$3 and 6 keV, respectively, and both host
at $z\sim0$ a prominent, central cD galaxy.

Comparing the simulations without and with baryons, it is found that
for the latter,
the inner DM density profiles, $r\la$ 50-100 kpc, steepen considerably:
$\Delta\alpha\sim$0.5-0.6, where -$\alpha$ is the logarithmic DM density
gradient. This is primarily due to the central stellar cDs becoming very 
massive, as a consequence of the onset of late time cooling flows and 
related star formation. Once these spurious cooling flows have been
corrected for, and the cluster gravitational
potentials dynamically adjusted, much smaller pinching effects are found:
$\Delta\alpha\sim$0.1. 

The main result of the paper is hence that
our work strongly supports the notion that the steepening of the
inner dark matter density profile, due to baryonic pinching, is very
moderate. This finding is at stark variance with findings for individual
field galaxies. The main reason for this is that galaxy clusters,
even in the presence of central dominant cD galaxies, are dark matter
dominated outside of $\sim$7-8 kpc. As the effective radii of the two
cDs considered are 11 and 16 kpc, this result indicates that cD galaxies
are dark matter dominated at the effective radius.

Including the effects of baryonic pinching,
central slopes of $\alpha\simeq$1.0 and 1.2 are found for the DM in the 
two clusters, interestingly close to recent observational findings for Abell~1703
based on gravitational lensing studies.

For the simulations with baryons, the inner density profile of DM+ICM
combined is found to be only very marginally steeper than that of the
DM. The total
matter inner density profiles, however, are found to be 
$\Delta\alpha\sim0.5$ steeper 
than the inner profiles in the dark matter only simulations. 
The total matter density profiles are not well described by 
generalized NFW models.

Given that we have presented results for just two simulated clusters
in this paper, it is clearly of importance to increase the sample size.
On the observational side, strong lensing constraints for more uni-modal 
clusters should be obtained.
We plan to present such results in forthcoming papers.

\section*{Acknowledgments}
We have benefited considerably from discussions with Sune Toft and 
Signe Riemer-S{\o}rensen.
We gratefully acknowledge abundant access to the computing facilities
provided by the Danish Centre for Scientific Computing (DCSC). This
work was supported by the DFG Cluster of Excellence ``Origin and Structure
of the Universe''. The Dark Cosmology Centre is funded by the Danish
National Research Foundation.
ML acknowledges the Centre National d'Etudes Spatiales (CNES) for their support.

\label{lastpage}

\end{document}